# Hopfion Dynamics in Chiral Magnets

Zulfidin Khodzhaev and Emrah Turgut

*Department of Physics, Oklahoma State University, Stillwater, Oklahoma 74078, USA*

## Abstract

Resonant spin dynamics of topological spin textures are correlated with their topological nature, which can be employed to understand this nature. In this study, we present resonant spin dynamics of three-dimensional topological spin texture, i.e., Neel and Bloch hopfions. Using micromagnetic simulations, we stabilize Bloch and Neel hopfions with bulk and interfacial Dzyaloshinskii-Moriya interaction (DMI), respectively. We identify the ground state spin configuration of both hopfions, effects of anisotropies, geometric confinements, and demagnetizing fields. To confirm topological nature, Hopf number is calculated for each spin texture. Then, we calculate the resonance frequencies and spin-wave modes of spin precessions under multiple magnetic fields. Unique resonance frequencies and specific magnetic field dependence can help to guide experimental studies to identify the three-dimensional topological spin texture of hopfions in functioning chiral magnets when imaging is not possible.

## I. INTRODUCTION

Topological spin textures attract considerable interest in condensed matter physics because of their non-trivial physical properties and promising applications in memory and storage technologies [1,2]. One well-known example of these spin textures is a magnetic skyrmion, which is a projection of three-dimensional hedgehog spins from a sphere to a two-dimensional surface [3,4]. When these three-dimensional hedgehog spins are projected on a torus with certain knotting, we end up having a hopfion, which is a three-dimensional topological spin texture. Although skyrmions have been intensely studied by theoretical, numerical, and experimental means [5–9], hopfions haven't been studied thoroughly and started to receive attention very recently [10–12].

Hopfions have been initially observed in chiral liquid crystals and optical mediums [12–14]. More recently numerical studies of the magnetic version of hopfions [11,15–17] and experimental observation in chiral magnet Ir/Co/Pt multilayers have been reported [18]. Magnetic hopfions are stabilized in chiral magnets as a result of antisymmetric exchange interaction, commonly called the DMI. Depending on the origin of DMI, i.e., bulk or interfacial DMI, Bloch or Neel hopfions are stabilized in magnetic nanodisks. The numerical studies of Bloch hopfions use interfaces with a strong perpendicular magnetic anisotropy (PMA) in addition to the bulk DMI in a B20 FeGe chiral magnet. A recent experimental study by Kent, et. al., used Ir/Co/Pt multilayer nanodiscs to stabilize magnetic hopfions and X-ray photoelectron emission microscopy (X-PEEM) and magnetic transmission X-ray microscopy (MTXM) to probe their internal spin texture [18]. For X-ray transmission, Ir/Co/Pt multilayer nanodiscs were fabricated on $Si_3N_4$ membranes. Besides, observation of Bloch hopfions in B20 compounds with bulk DMI stays a challenge due to epitaxial growth of B20 compounds on $Si_3N_4$ membranes. The requirement of $Si_3N_4$ membrane is also present in other commonly used methods of Lorentz transmission electron microscopy (TEM) imaging [19]. Scalable growth and device fabrication of hopfion systems require more stable substrates (e.g., Si), which will, however, make imaging studies more complicated.

Another important method of exploring nanoscale magnetic materials is studying their resonance spin dynamics [20–23]. For example, ferromagnetic resonance spectroscopy (FMR) has been used to characterize effective damping, magnetization, and other magnetic interactions in ferromagnetic materials [24,25]. Resonance dynamics in more complex spin textures, such as chiral magnets and magnetic skyrmions [26–30], can reveal further information about the real-space spin texture and spin-wave modes, in addition to the information obtained from ferromagnetic resonance spectroscopy aforementioned. Moreover, understanding the spin dynamics would be crucial for some applications, such as auto-oscillators made of chiral magnets [31].

In this article, we numerically study resonance spin precession dynamics of Bloch and Neel hopfions in chiral magnet nanodiscs utilizing micromagnetic simulations. We first introduce the geometry of the hopfion and hopfion charge. Then, we identify the ground state of the Bloch hopfion in the chiral magnet FeGe with bulk DMI and the Neel hopfion in magnetic multilayer with interfacial DMI. Particularly, we investigated the effect of the demagnetizing field in FeGe nanodisks, which was previously ignored in some studies [11,31]. We found that the demagnetization field stabilizes a Bloch hopfion in small discs against the monopole-antimonopole pair (MAP) formation, which is another stable spin configuration due to strong PMA. Next, we excite the Bloch and Neel hopfions with a magnetic pulse and analyze the resulting precession frequencies and spin-wave modes, which then can be correlated with the hopfion's real-space spin texture. Our results will provide the precession dynamics of three-dimensional spin-solitons and help their future experimental studies.

A hopfion is a knot on a three-dimensional torus consisting of continuous unit vector fields. In FIG. 1, we show three different hopfions with hopfion charges of 1, 2, and 3. The mathematical description of a hopfion charge is:

$$H = -\frac{1}{(8\pi)^2} \int \boldsymbol{F} \cdot \boldsymbol{A} \, d\boldsymbol{r}, \tag{1}$$

where $\boldsymbol{F}$ and $\boldsymbol{A}$ are vector fields calculated from the unit vector field $\boldsymbol{n}$ by:

$$F_i = \epsilon_{ijk} \boldsymbol{n} \cdot (\nabla_j \boldsymbol{n} \times \nabla_k \boldsymbol{n}) \text{ and } \nabla \times \boldsymbol{A} = \boldsymbol{F}. \tag{2}$$

In our magnetic system, $\boldsymbol{n}$ corresponds to the unit microspin. Moreover, there is an indirect method to calculate the hopfion charge using the linking number of knots [32,33]; as we demonstrate in FIG. 1, where we describe the torus by its toroidal (*T*) and poloidal (*P*) cycles and the hopfion charge $H = TP$ [32,33]. The left column of FIG. 1 shows these isosurfaces with a color-coding, whereas the right column shows one isosurface with a knot that is the point with the magnetization vector angles of $\theta_i$ and $\phi_i$ on a unit sphere. Isosurfaces knot the torus one, two, and three times in FIGs. 1(a), 1(b), and 1(c), respectively.

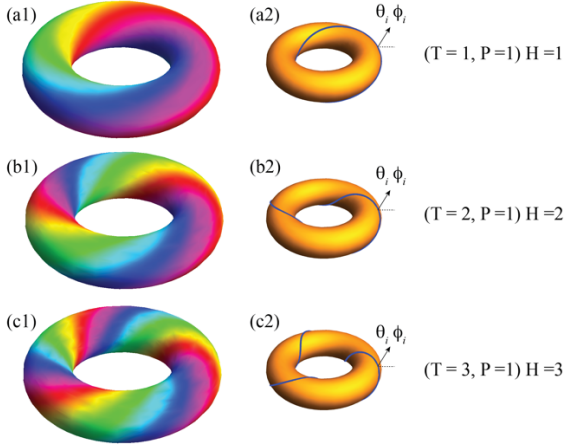

FIG. 1. *Illustration of hopfions with different toroidal (T) and poloidal (P) cycles; (a1), (b1), and (c1) show the isosurfaces with color-coding for hopfion charges one, two, and three, respectively. The lines in (a2), (b2), and (c2) show the knots of the coordinate $\theta_i$ and $\phi_i$ on a corresponding sphere.*

Hopfions being as spin textures in condensed matter systems have been previously introduced and explored theoretically as more abstract phenomena [34]. However, recent numerical studies have shown the stability of Hopfions in the chiral magnets with the help of DMI [11,15,35]. There are two origins of DMI in chiral magnets, the bulk DMI from the broken crystal inversion symmetry (e.g., noncentrosymmetric B20 FeGe [26,29,36]) and the interfacial DMI from the broken interface symmetry (e.g., Ir/Co/Pt multilayers [18,37]). The former facilitates Bloch hopfions and the latter facilitates Neel hopfions, similar to Bloch and Neel skyrmions [38]. Their spin profiles and dynamics are quite different; therefore, we first investigate Bloch hopfion in a FeGe nanodisk and then Neel hopfion in a magnetic multilayer.

## II. BLOCH HOPFION

For a Bloch hopfion, we use the initial ansatz and nanodisc geometry from Ref [11] to find its ground state using MuMax3 micromagnetic simulation [39]. Our simulation consists of a FeGe nanodisc with a diameter of 128 nm and a height of 64 nm. We sandwiched the FeGe layer by two PMA layers with an anisotropy constant of $K_u = 10^5$ J/m$^3$, which can be experimentally created by oxide or heavy metal layers [40]. To find the ground state, we minimized the total energy of the system, including exchange, DMI, anisotropy, and demagnetizing energies by using micromagnetic simulation. Magnetic interactions and properties of FeGe are highly sensitive to temperature and we use these values near the skyrmion formation temperature of 276 K [29]. Besides, the skyrmion formation temperature in the FeGe is the highest among all the B20 components with a skyrmion phase, which makes the FeGe more promising for technological applications [38].

One surprising difference we observed in our simulations is that a hopfion in a nanodisc with a 128 nm diameter is stable, even though Tai *et al.*[11] found that the hopfion phase can only be stable in larger nanodiscs, i.e., with a diameter of $D > 3\lambda$, where $\lambda = 70$ nm (the helical period of FeGe). The difference between our micromagnetic simulation and Tai *et al.* [11]'s numerical method is that we account for the demagnetizing energy, which originates from the shape of the structure. In FIG. 2, we show the resulting spin textures by including and excluding the demagnetizing field in the total energy at zero external magnetic fields. First, we start the simulation with the initial ansatz of a hopfion [11,34], as in FIG. 2(a) and FIG. 2(b), which shows the magnetization of the disc at $y = 64$ nm and $z = 32$ nm planes, respectively. The color coding for the magnetization maps is given in FIG. 2(i). When we exclude the demagnetizing field as in Tai *et al.* [11], the MAP ground state is reached [FIG. 2(c) and Fig. 2(d)], which is consistent with their observation. However, when the demagnetization field is included, the hopfion becomes a stable state as shown in FIG. 2(e) and FIG. 2(f). In addition, we calculated the exchange (symmetric + DMI) and demagnetization energy densities in the case of the hopfion state [FIG. 2(g) and FIG. 2(h)]. Although the demagnetizing energy is quite smaller than the sum of the two exchange energies, it is still in the same order; and especially, it is strong at the center and at the edges of the disc, where the spins are aligned along +z direction. Therefore, we found that the demagnetizing field plays an important role and increases the stability of the hopfion phase, especially, in smaller diameter discs. This is particularly important because the smaller discs with a stable hopfion phase means higher density for memory applications.

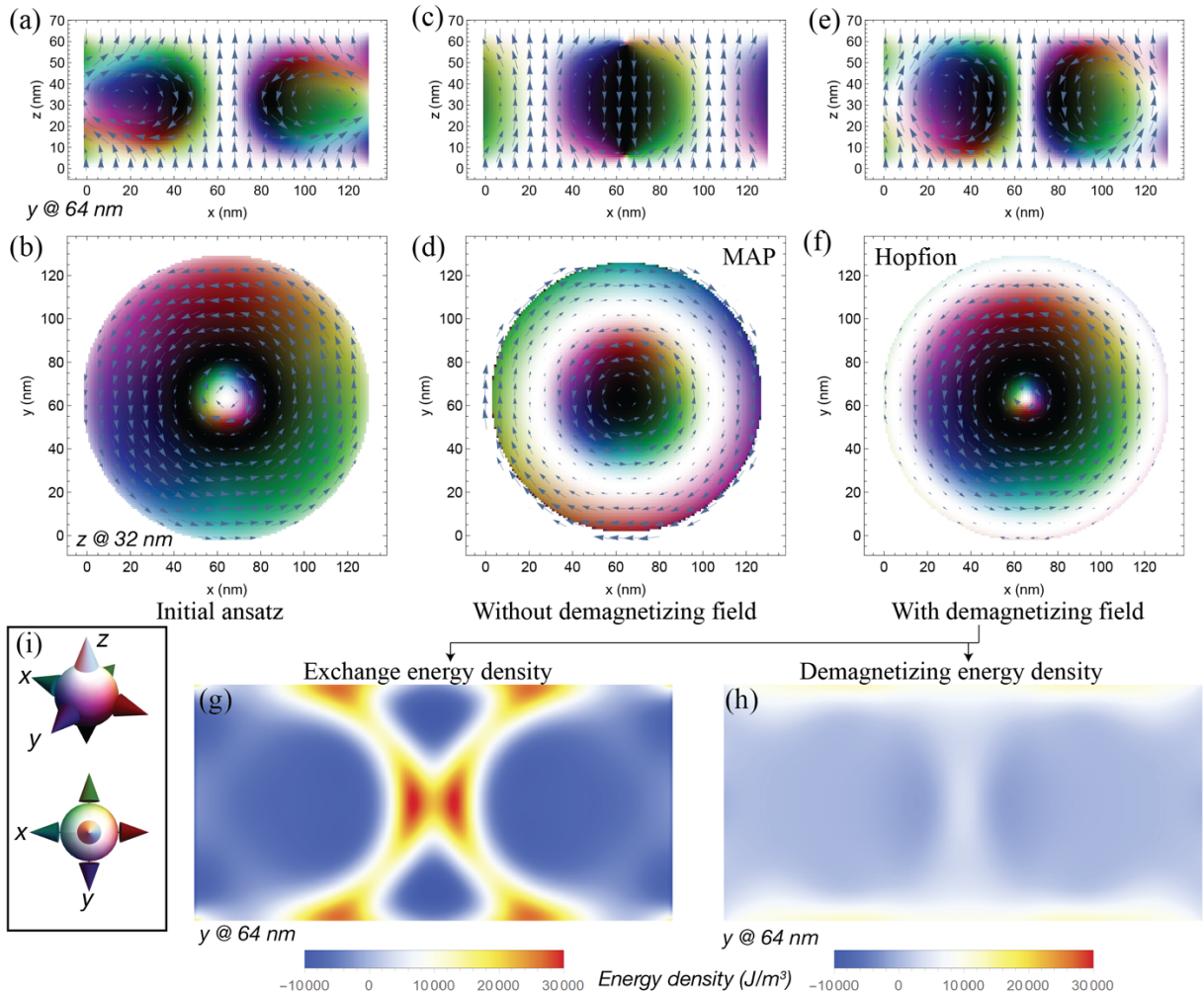

FIG. 2. Ground spin textures with and without the demagnetizing field. The simulation disc has 128 nm diameter and 64 nm height. (a), (c), and (e) are the magnetizations at y = 64 nm plane. (b), (d), and (f) are the magnetizations at z = 32 nm plane. (a) and (b) are the initial magnetization configuration to start the energy minimization. (c)-(d) and (e)-(f) are the MAP and hopfion ground-state spin textures without and with the demagnetizing field, respectively. (i) shows the color-coding for the magnetization. (g) and (h) show the exchange energies and demagnetizing energy densities at the y =64 nm plane.

After finding the hopfion ground state in the FeGe nanodisc at zero magnetic fields, we study the stability of the hopfion phase at the external magnetic fields. A recent study by Liu et al. [16] already discussed the effects of external fields; therefore, we leave the effects of the external fields to Ref[16] and focus on its spin dynamics under various external magnetic fields. In our simulations, the hopfion is stable between 0 mT and -25 mT external magnetic fields in the direction of z. At higher magnetic fields, the hopfion state turns into the MAP state and the field polarized state is the ground state at lower magnetic fields.

To find the resonance frequencies, we employ the ringdown method [41], in which we apply a magnetic field impulse and record the magnetization of each unit cell with a $\Delta t = 0.1$ ns time interval for $\tau = 200$ ns. By calculating the Fourier transform of the magnetization in the time domain, we find the spatially resolved spin dynamics in the frequency domain with a 5 MHz resolution up to 5 GHz in frequency. The averaged Fourier transforms of all spins are shown in FIG. 3 for $B$ = -5, -10, -15, -20, and -25 mT magnetic fields. For clarity, the y positions of the spectrums are shifted by an offset.

As the magnetic field amplitude increased from -5 mT to -25 mT, the first resonance frequency $f_1$ at 0.11 GHz shifts to higher frequencies, e.g., $f_1$ = 0.14, 0.17, 0.19, and 0.20 GHz at B = -10, -15, -20, and -25 mT fields, respectively. Another strong resonance feature consistent with the varying field is around 1.75 GHz, which does not exhibit a monotonic increase. At the field range from -5 mT to -25 mT, this resonance is located at $f_2$ = 1.82, 1.78, 1.75, 1.73, and 1.74 GHz. Next, we investigate the spin-wave modes of $f_1$ and $f_2$ resonances by plotting spatial distributions of the magnetization (Fourier) components in $z$, $\rho$, and $\phi$ cylindrical components at $B$ = -15 mT field (FIG. 4). We show the spin-wave modes in z = 32 nm (first two rows of FIG. 4) and z = 16 nm (third and fourth rows of FIG. 4) of micromagnetic simulations, which are the middle plane and one-fourth plane of the torus, respectively. We notice that the middle plane (z = 32 nm) has mainly spin waves close to the center of the hopfion, while z = 16 nm shows strong spin waves distributed around the hopfion. The resonance at 0.17 GHz has stronger contributions from z = 32 nm, and other layers start to contribute less, while there is a small change in 1.75 GHz resonance between z = 32 nm and z = 16 nm. Although our average spectrums from FIG. 3 shows multiple resonances, our spin-wave analysis indicates that 0.17 and 1.75 GHz resonances are consistent throughout the five magnetic fields.

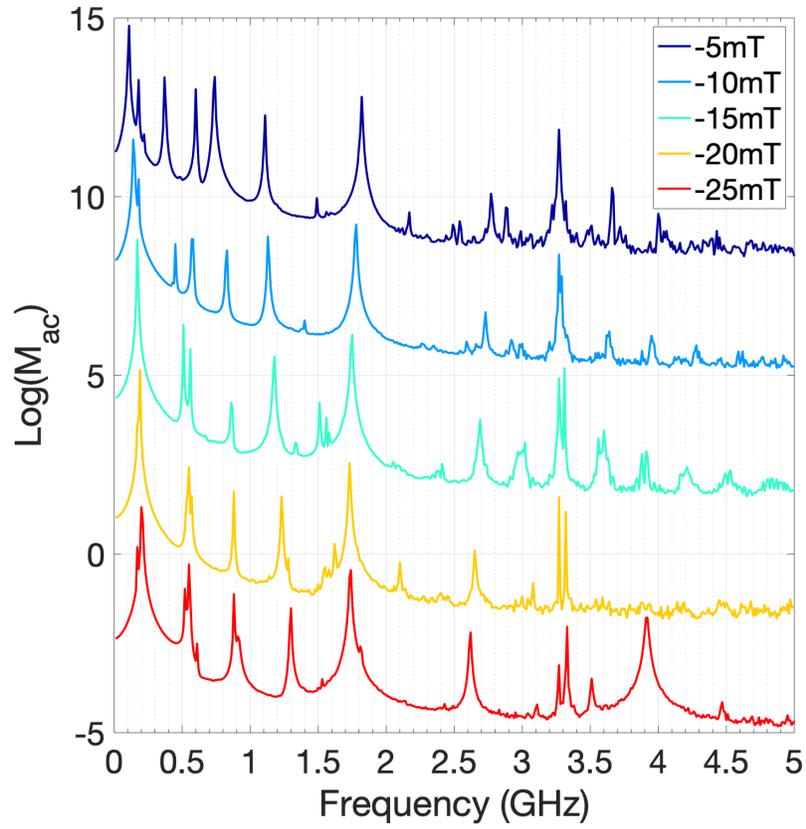

FIG. 3. Resonance frequencies of the hopfion at B = -5, -10, -15, -20, and -25 mT fields in the direction of z-axis. The curves are the sum of the Fourier transform of all the individual spin dynamics. For clarity, the spectrums are shifted by an offset.

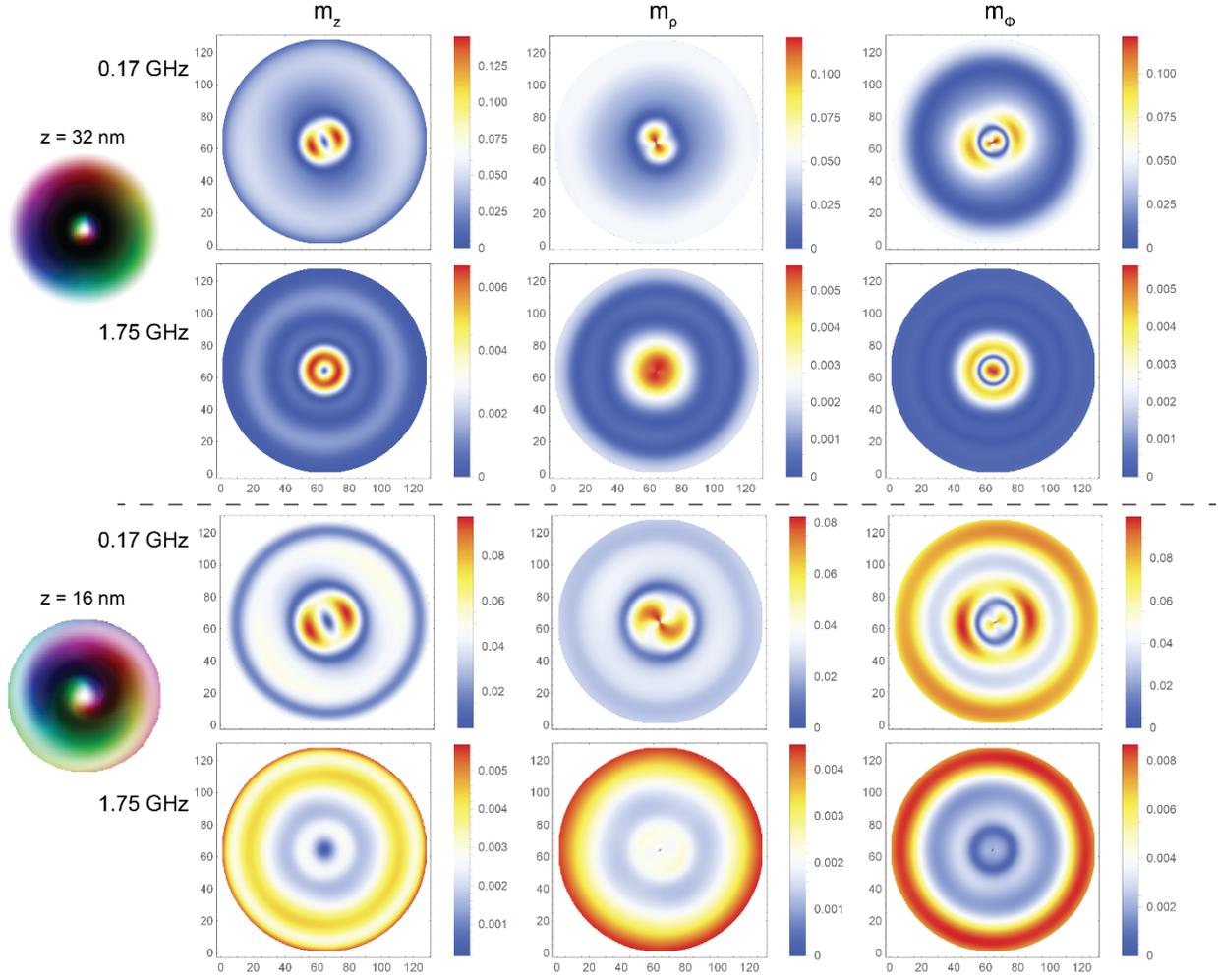

FIG. 4. Spatial distribution of the resonance amplitudes at 0.17 GHz (first and third rows) and 1.75 GHz (second and fourth rows) frequencies at -15 mT magnetic field. The top two rows show the amplitude at the middle plane of the hopfion at z = 32 nm and the bottom two rows show at z = 16 nm. The columns show the amplitude for the magnetization in z, $\rho$, and $\phi$ directions, respectively.

## III.   NEEL HOPFION

Neel hopfion is also a knot on a three-dimensional torus consisting of continuous unit vector fields with the magnetization vector aligned along with $\rho$ direction on a unit sphere whereas the Bloch hopfion's magnetization vector is aligned along with $\phi$ direction in the cylindrical coordinates. Previously, Neel hopfions were numerically [42] and experimentally [18] studied in systems with interfacial DMI.

We rotate the individual spins from a Bloch hopfion's initial ansatz [40] ninety degrees around the z-axis, i.e., rotating magnetization along the axis with masking coordinates as $x \to y$ and $y \to -x$. By using MuMax3 micromagnetic simulation, we found the ground state of a Neel hopfion as shown in FIG 5. The magnetic energy parameters used in the simulations are the interfacial DMI

= 1.15×10⁻³ J/m², the exchange stiffness $A_{ex}$ = 1.1×10⁻¹² J/m, the uniaxial anisotropy constant $K_u$ = 1×10³ J/m³, and the saturation magnetization $M_{sat}$ = 3×10⁵ A/m. The simulation size for the Neel hopfion is a disk with a diameter of 64 nm and a height of 8 nm.

The hopfion charge for the Bloch hopfion is already extensively studied in [11,16]; however, Neel hopfions are relatively new and we will confirm the topological nature of our Neel hopfions. The topological charge is calculated using Ref [42]'s method as follows:

$$H = \frac{1}{(2\pi)^2} \iiint_{x,y,z}(\mathbf{F_x} \cdot \mathbf{A_x} + \mathbf{F_z} \cdot \mathbf{A_z})\, dxdydz, \tag{3}$$

$$\mathbf{A_x} = -\int_{-\infty}^{y} F_z\, dy, \tag{4}$$

$$\mathbf{A_z} = \int_{-\infty}^{y} F_x\, dy, \tag{5}$$

where $F_i = \epsilon_{ijk}\mathbf{n} \cdot (\nabla_j \mathbf{n} \times \nabla_k \mathbf{n})$. We found that the Neel hopfion is stable in the field range of -30 mT to -440 mT, where the Hopf indices $H$ are above 0.9, indicating that all solitons are Neel hopfions. We show the magnetizations of hopfions at the fields of -60 mT, -120 mT, and -300 mT fields in FIG. 5, where the slices of $xy$ plane at z = 4 nm and $xz$ plane at y = 32 nm are shown. The color coding for the magnetization maps is as in FIG. 2(i). As the external field decreased to -300 mT, the size of the hopfion increases and its shape starts to change from a circular to a square. After we found the stable magnetic field range, we studied resonance dynamics from -30 to -300 mT magnetic field range.

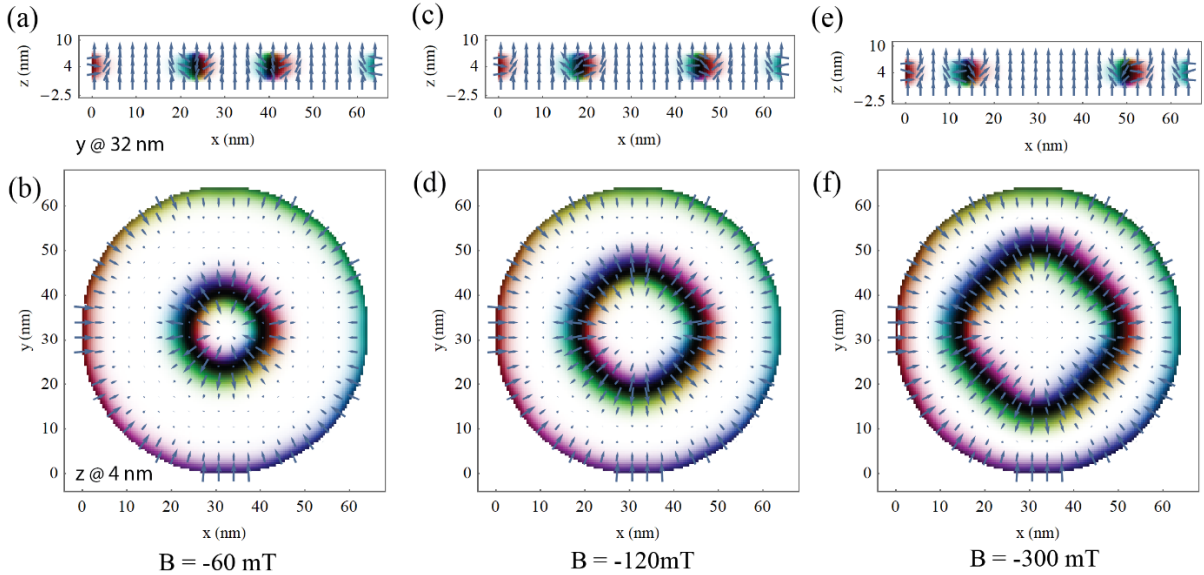

FIG. 5. Neel hopfion with B = -60 mT, -120 mT, and -300 mT fields. The simulation disc has 64 nm diameter and 8 nm height. (b), (d), and (f) are the magnetizations at z = 4 nm plane and (a), (c), and (e) are the magnetizations at y = 32 nm plane under demagnetizing fields of -60 mT, -120 mT, -300 mT, respectively.

Neel hopfion dynamics are studied as in the previous section. We apply a magnetic field impulse and the magnetization of each unit cell is recorded with a $\Delta t = 0.1$ ns time interval for $\tau = 200$ ns. Using the Fourier transform of the magnetization in the time domain, spatially resolved spin dynamics are calculated in the frequency domain with a 5 MHz resolution up to 5 GHz. The averaged Fourier transforms of all spins are shown in FIG. 6 for B = -30, -60, -90, -120, -150, -180, -210, -240, -270, and -300 mT magnetic fields. For clarity, the y positions of the spectrums are shifted by an offset.

As the magnetic field amplitude increases from -30 mT to -60 mT, the first resonance frequency $f_1$ at 3.2 GHz shifts to lower frequencies, e.g., f = 2.19 and 2.13 GHz at B = -60 and -90 mT fields, respectively. After 2.13 GHz, resonance frequencies shift to high frequencies, i.e., f = 2.49, 2.9, 3.65, 3.98, 4.27 and 4.51 GHz at B = -120, -150, 180, 210, 240, 270 and -300 mT fields, respectively. After, we investigate the spin-wave modes of 1.26 GHz, 2.19 GHz, and 4.39 GHz resonances by plotting spatial distributions of the magnetization components in z, $\rho$, and $\phi$ cylindrical coordinates at B = -60 mT field (FIG. 7). We show the spin-wave modes in z = 4 nm xy plane (first three rows of FIG. 7) and z = 2 nm (fourth, fifth and sixth rows of FIG. 7) of micromagnetic simulations, which are the middle plane and one-fourth plane of the torus, respectively.

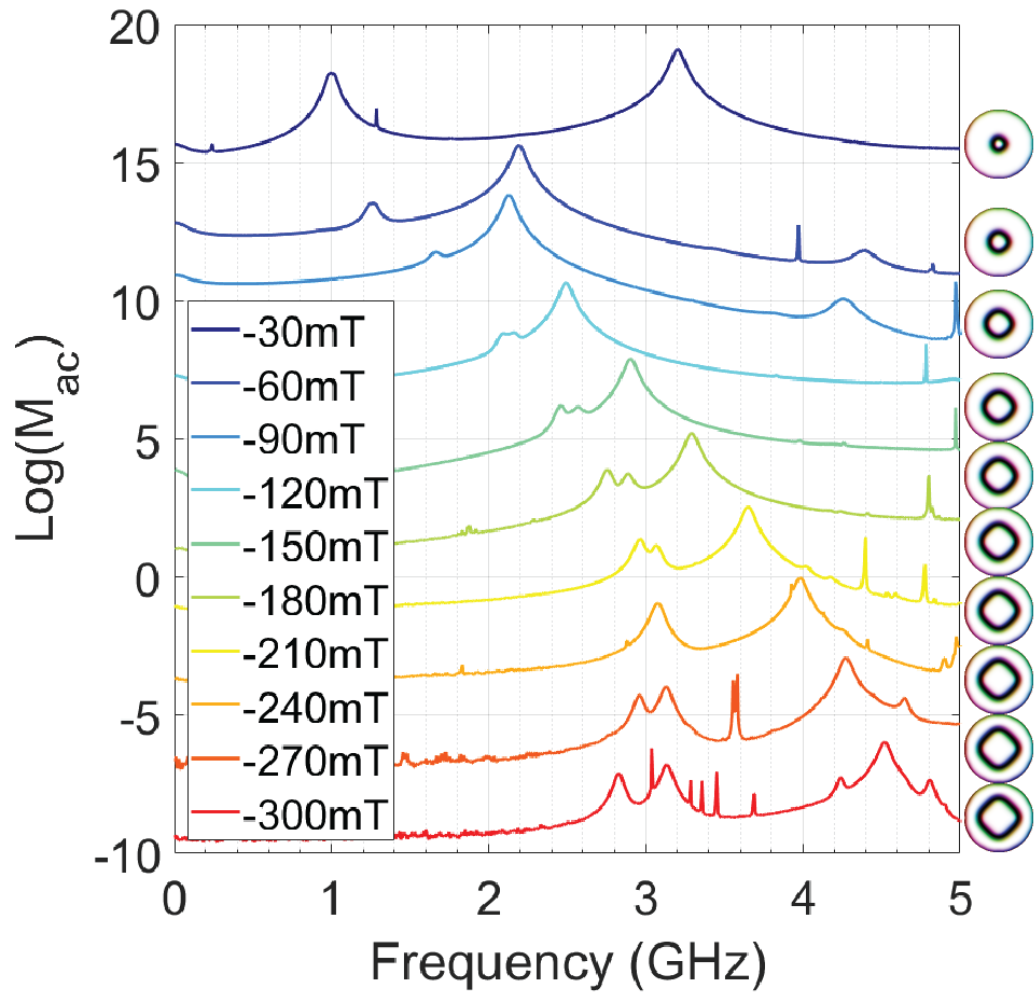

FIG. 6. Resonance frequencies of the Neel hopfion at B = -30, -60, -90, -120, -150, -180, -210, -240, -270 and -300 mT fields in the direction of z-axis. The curves are the sum of the Fourier transform of all the individual spin dynamics. For clarity, the spectrums are shifted by an offset.

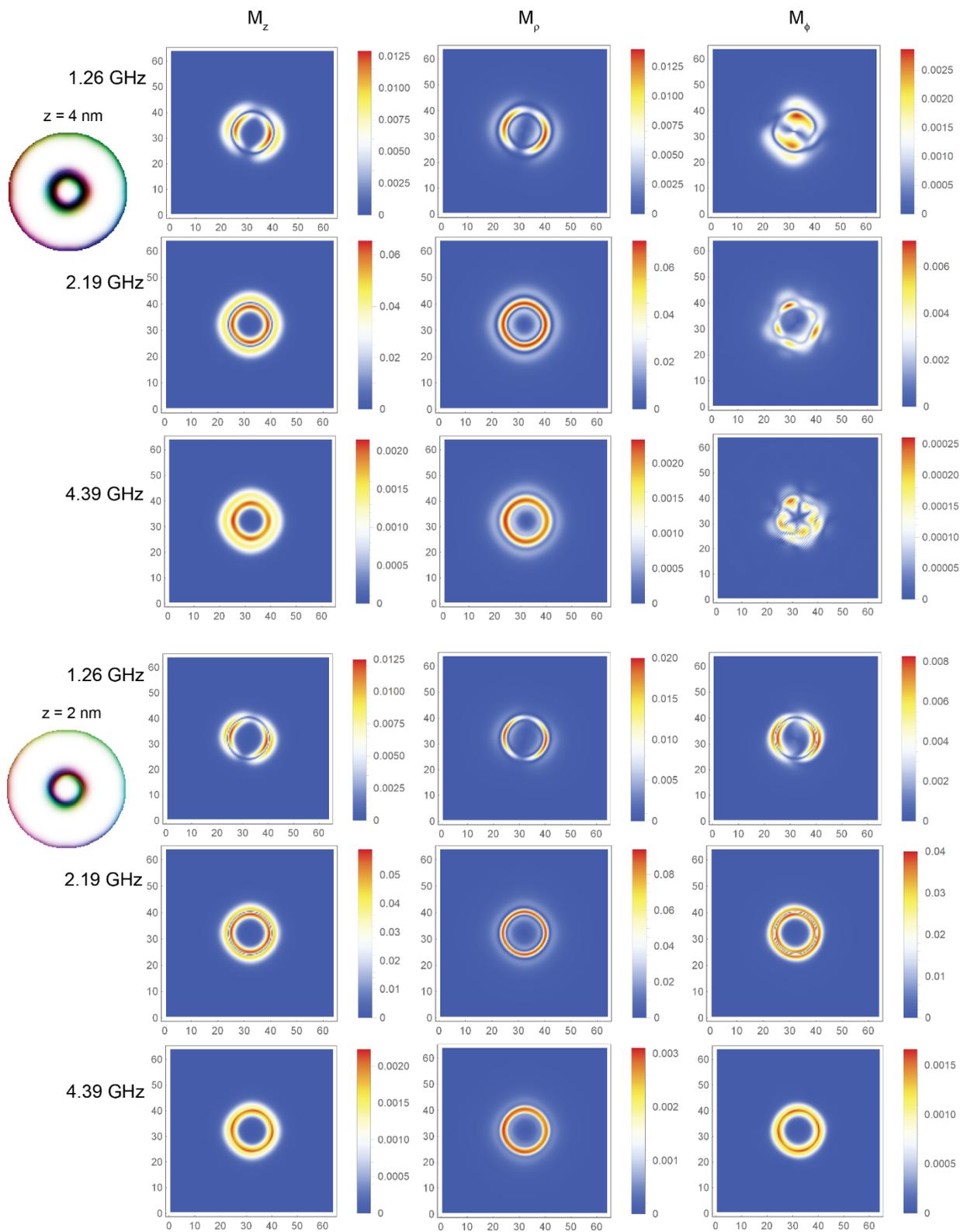

FIG. 7. Spatial distribution of the resonance at -60 mT magnetic field. The top three rows show the amplitude at the middle plane of the hopfion at z = 4 nm xy plane and the bottom three rows show at z = 2 nm.

In summary, we numerically study the stability and spin dynamics of a three-dimensional spin texture of Bloch and Neel hopfions using micromagnetic simulations. Bloch hopfion's stability under different external magnetic fields in the z-direction, using 128 nm nanodisc, was analyzed. At lower magnetic fields, the Bloch hopfion turned into a field polarized state and at higher fields, into a MAP. We found that the demagnetizing field helps to stabilize Bloch hopfion with two times smaller disks, which is useful for higher density memory devices. By analyzing the transient dynamics of the Bloch hopfion spins, we found the resonance features, spin-wave modes, and the spatial distribution of spin dynamics at two different frequencies. In the second part of the paper, Neel hopfion is discussed. The stability of Neel hopfion was achieved under multiple magnetic fields pointing in the z-direction, using a 64 nm diameter disk. At higher external magnetic fields, hopfion's shape turned into a square. Then, resonance and spin waves were analyzed for varying external magnetic fields. Lastly, the spatial distribution of spin dynamics at three different frequencies is shown. Further investigation by varying thermal fluctuations and anisotropies by electric fields will help to further manipulate resonance features. These will lead to identifying the true nature of the spin texture when magnetic imaging or neutron scattering on functioning devices are not possible.

## Acknowledgment

This work was supported by the National Science Foundation grant number OIA-1929086. We also gratefully acknowledge NVIDIA Corporation with the donation of the Titan Xp GPU, which was used for this research.